\documentclass[12pt,a4]{article} 
\newcommand{\vs}{\vspace{-0.25cm}}
\usepackage{amsmath,graphicx}
\begin{document} 
\begin{center}
  {\Large{\bf Density-dependent NN-interaction from subleading chiral
3N-forces: short-range terms and relativistic corrections}\footnote{This work 
has been supported in part by DFG and NSFC (CRC110).}  }  

\medskip

 N. Kaiser and V. Niessner \\
\medskip
{\small Physik-Department T39, Technische Universit\"{a}t M\"{u}nchen,
   D-85747 Garching, Germany\\

\smallskip

{\it email: nkaiser@ph.tum.de}}
\end{center}
\medskip
\begin{abstract}
We derive from the subleading contributions to the chiral three-nucleon force
(short-range terms and relativistic corrections, published in Phys. Rev. C84, 054001 (2011))
a density-dependent two-nucleon interaction
$V_\text{med}$ in isospin-symmetric nuclear matter. The momentum and
$k_f$-dependent potentials associated with the isospin operators ($1$ and $
\vec\tau_1\!\cdot\!\vec\tau_2$) and five independent spin-structures are
expressed in terms of loop functions, which are either given in closed
analytical form or require at most one numerical integration. Our results for
$V_\text{med}$ are most helpful to implement subleading chiral 3N-forces into
nuclear many-body calculations.
\end{abstract}

\section{Introduction and summary}
Three-nucleon forces are an indispensable ingredient in accurate few-nucleon and
nuclear structure calculations. Nowadays, chiral effective field theory is the
appropriate tool to construct systematically the nuclear interactions \cite{evgeni}. Precise
two-nucleon potentials have been developed at next-to-next-to-next-to-leading order
(N$^3$LO) in the chiral (small momentum) expansion \cite{evgeni,hammer,machleidtreview}.
The extension of the NN-potential to N$^4$LO, with
various higher-order two- and three-pion exchange contributions, has been accomplished in
refs.\,\cite{n4lo1,n4lo2}. Three-nucleon forces appear first at N$^2$LO, where they
consist of a zero-range contact-term (parameter $c_E$), a mid-range $1\pi$-exchange
component (parameter $c_D$) and a long-range $2\pi$-exchange component (parameters
$c_{1,3,4}$). The calculation of the subleading chiral three-nucleon forces, built up
by many pion-loop diagrams, has been performed for the long-range contributions in
ref.\cite{3Nlong} and completed with the short-range terms and relativistic corrections
in ref.\,\cite{3Nshort}. Moreover, the extension of the $2\pi$-exchange component of the
chiral three-nucleon force to N$^4$LO has been acomplished in ref.\,\cite{3Nfourth} and the
corresponding (ten independent) 3N-contact terms quadratic in momenta have been derived
in ref.\,\cite{3Ncontact}.

For the variety of existing many-body methods that are commonly employed in calculations of
nuclear matter or medium mass and heavy nuclei it is technically very challenging
to include chiral three-nucleon forces directly. In ref.\,\cite{hebeler} a decomposition
of the chiral three-nucleon forces at N$^2$LO and N$^3$LO in a momentum-space partial wave basis
has been proposed, which should make ab initio studies of few-nucleon scattering, nuclei
and nuclear matter computationally much more efficient. In that endeavor one is working with
large file sizes for the 3-body matrix elements in each individual spin-isospin channel, while
the values of the low-energy constants and the form of the regulator function can be chosen
freely. An alternative and simpler approach is to employ instead a density-dependent
two-nucleon interaction $V_\text{med}$ that reflects the underlying three-nucleon force. The
analytical calculation of $V_\text{med}$ from the leading chiral 3N-force at N$^2$LO (involving
the parameters $c_{1,3,4}$, $c_D$ and $c_E$) has been presented in ref.\,\cite{holt}.
When restricting to on-shell scattering of two-nucleons in isospin-symmetric spin-saturated
nuclear matter, the resulting in-medium NN-interaction  $V_\text{med}$ has the same isospin-
and spin-structure as the free NN-potential. The subsequent decomposition into partial wave
matrix elements has provided a good illustration of the (repulsive or attractive) effects of the
various components of $V_\text{med}$ in different spin-isospin channels. Together with extensions to isospin-asymmetric nuclear matter or spin-polarized neutron matter, the in-medium interaction $V_\text{med}$ derived from the leading chiral 3N-force has found many applications in recent years \cite{corbinian1,corbinian2,spinpola}. Furthermore, the approach has been generalized to the strangeness sector by deriving in-medium hyperon-nucleon interactions \cite{YNN} from the leading order chiral YNN-forces (where Y=$\Lambda, \Sigma$).

The purpose of the present paper is to continue the construction of the in-medium
NN-interaction $V_\text{med}$ to N$^3$LO by treating the subleading contributions to the
chiral 3N-force. We focus here on the short-range terms and relativistic corrections, which have been derived in ref.\,\cite{3Nshort} by using the method of unitary transformations. The calculation of $V_\text{med}$ from the remaining long-range 3N-forces (divided into diagram classes of two-pion-exchange topology, two-pion-one-pion-exchange topology, and ring topology in ref.\,\cite{3Nlong}) is yet more demanding and relegated to a future publication. 
In section 2, we outline the diagrammatic calculation of $V_\text{med}$ from a generic 3N-interaction $V_\text{3N}$ by closing one nucleon-line to an in-medium loop. This way one encounters four different topologies of in-medium NN-scattering diagrams: self closings, short-range vertex corrections, pionic vertex corrections, and double exchanges. The basis of five independent spin-operators, into which the in-medium interaction $V_\text{med}$ is decomposed, is also given. In section 3 the one-pion-exchange-contact
part of the 3N-force is treated and shown to produce an effectively vanishing in-medium
NN-interaction $V_\text{med}^{(^1S_0)}=0$. Section 4 deals with the two-pion-exchange-contact
topology and analytical expressions are given for the obtained contributions to
$V_\text{med}$. In section 5 the leading relativistic corrections to the chiral 3N-force,
divided into $1\pi$-exchange-contact and $2\pi$-exchange topologies, are treated. The
corresponding contributions to $V_\text{med}$ are expressed in terms of loop-functions
$\Gamma_\nu(p,k_f), \gamma_\nu(p,k_f), G_\nu(p,q,k_f)$ and $K_\nu(p,q,k_f)$ which are listed 
in their explicit form in the appendix. Moreover, in section 6 the subleading 
3N-contact potential of ref.\,\cite{3Ncontact} is converted into an in-medium NN-interaction 
$V_\text{med}$ linear in density $\rho=2k_f^3/3\pi^2$ and quadratic in momenta.

In summary, after eventual partial-wave projection (see herefore e.g. eqs.(38)-(41) in
ref.\cite{holt}) our results for $V_\text{med}$ are suitable for easy implementention of
subleading chiral 3N-forces into nuclear many-body calculations.

\section{In-medium NN-interaction from one-loop diagrams}
\begin{figure}[ht]\centering
\includegraphics[width=0.3\textwidth]{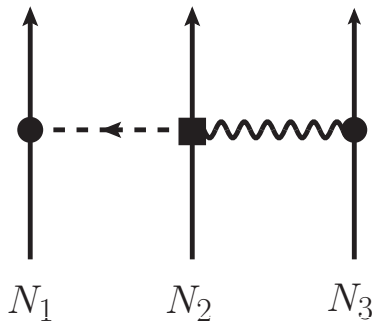}
\caption{Generic form of the 3N-interaction. The dashed line symbolizes 
pion-exchange and the wiggly line a short-range interaction.} \end{figure}

First, we outline how the in-medium interaction  $V_\text{med}$ is constructed and calculated
diagrammatically. The generic form of a chiral 3N-interaction $V_\text{3N}$ is shown in
Fig.\,1. The dashed line symbolizes (one or two) pion-exchange and the wiggly line a short-range
interaction (without a momentum-dependent propagator). In order to obtain from $V_\text{3N}$ a
density-dependent NN-interaction, one has to close one nucleon-line and integrate the
resulting loop,  $(2\pi)^{-4}\!\int\!d^4l$, over the medium part $-2\pi\delta(l_0)\theta(k_f-|\vec l\,|)$ of the heavy nucleon particle-hole propagator. The process of closing one of the three identical fermionic nucleon-lines in
$V_\text{3N}$ leads to four different topologies for NN-scattering diagrams: self closings,
short-range vertex corrections, pionic vertex corrections, and double exchanges, which are
separately shown in Figs.\,2,\,3,\,4 and 5. The respective mirror graphs, resulting from
the interchange of both nucleons $N_1\leftrightarrow N_2$, have of course to be
added. With reference to these four topologies, the
contributions to the in-medium NN-interaction are distinguished and denoted as
$V_\text{med}^{(0)}, V_\text{med}^{(1)}, V_\text{med}^{(2)}$ and $V_\text{med}^{(3)}$. In the case of a $2\pi$-exchange 3N-force the two types of pionic vertex corrections will be combined into just one expression for $V_\text{med}^{(1)}$. This reduced specification is employed in subsections 5.2 and 5.3. Individual contributions to $V_\text{med}$ are represented in terms of loop functions, that are defined by Fermi sphere integrals $(2\pi)^{-1}\! \int\!d^3 l\,\theta(k_f-|\vec l\,|)$ over pion propagators. 

\begin{figure}[h]
\centering
\includegraphics[width=0.45\textwidth]{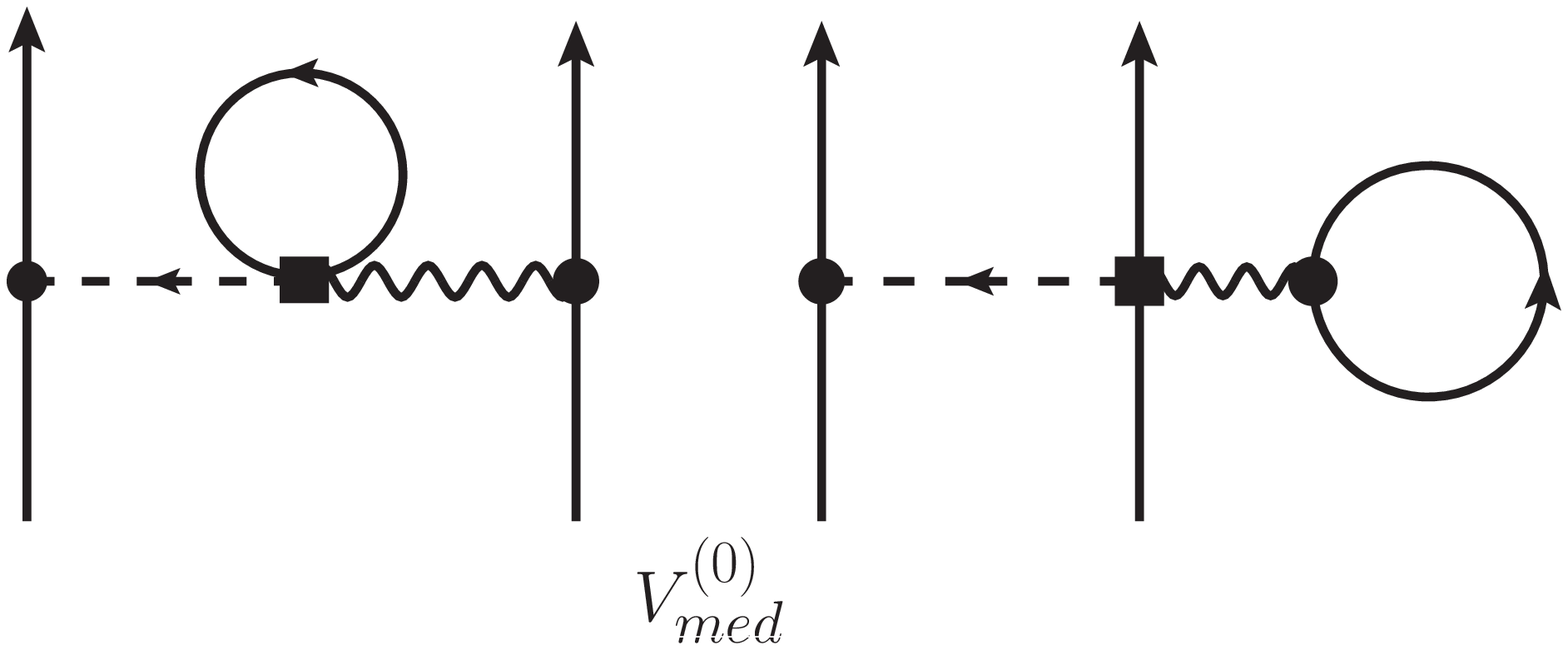}
\caption{Self closings of a nucleon line generating the contribution
$V_\text{med}^{(0)}$ to the in-medium NN-interaction. Mirror graphs $N_1
\leftrightarrow N_2$ have to be supplemented.} \vspace{0.5cm}
\centering
\includegraphics[width=0.45\textwidth]{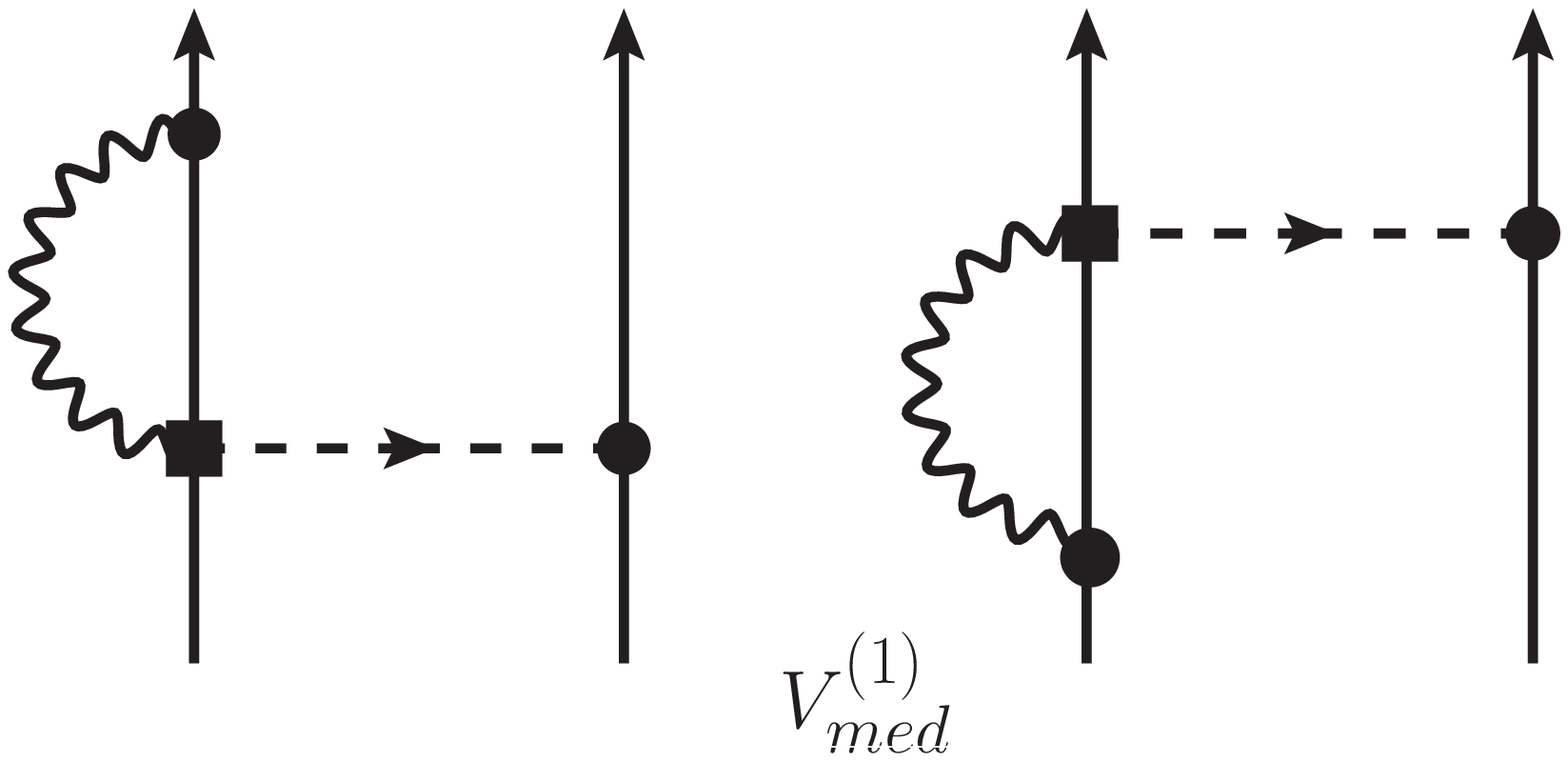}
\caption{Vertex corrections by the short-range interaction  generating the
contribution $V_\text{med}^{(1)}$. Mirror graphs $N_1
\leftrightarrow N_2$ have to be supplemented.}\vspace{0.5cm}
\centering
\includegraphics[width=0.45\textwidth]{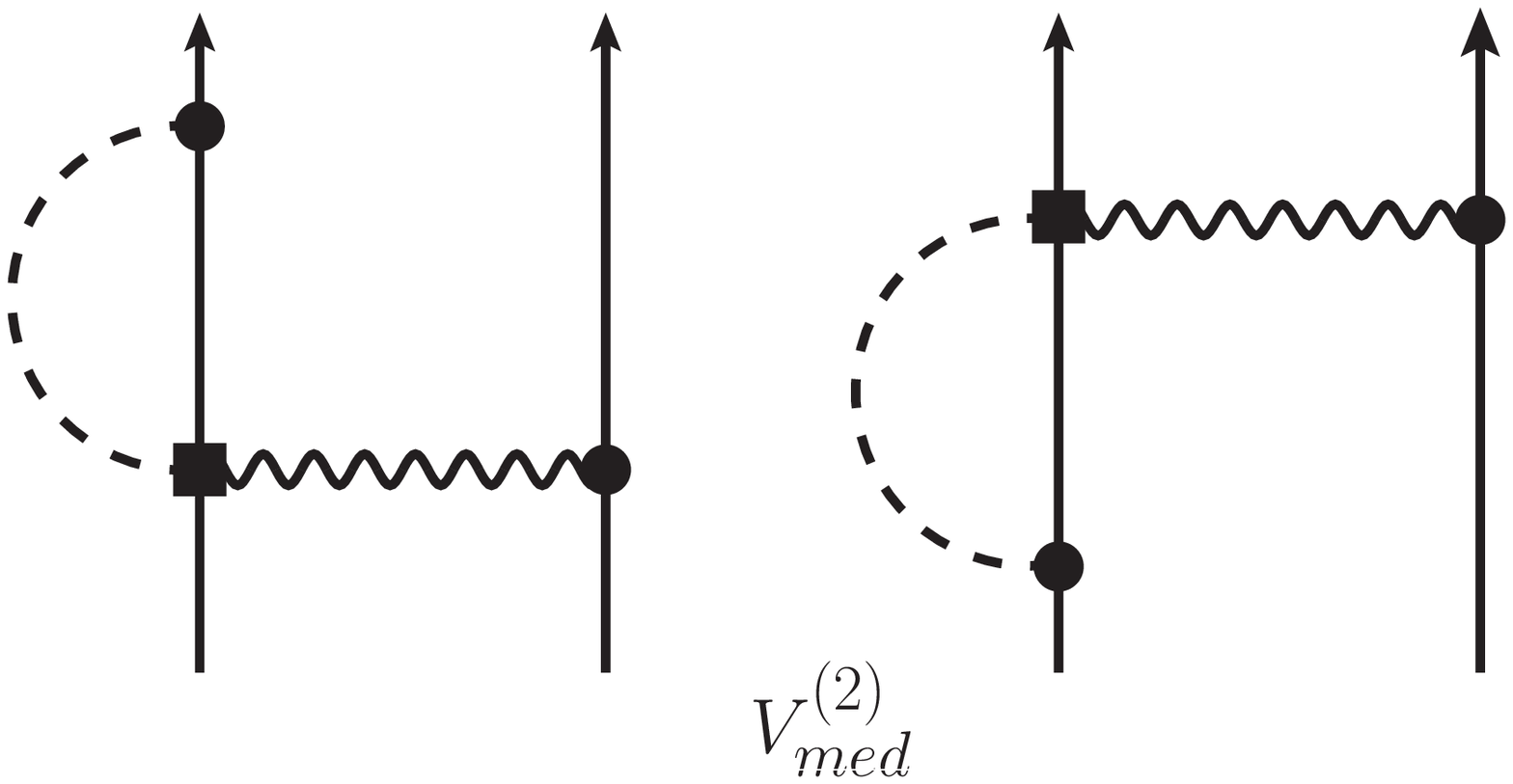}
\caption{Vertex corrections by pion-exchange generating the contribution
$V_\text{med}^{(2)}$. Mirror graphs $N_1\leftrightarrow N_2$ have to be
supplemented.} \vspace{0.5cm}\end{figure}
\begin{figure}[h]
\centering
\includegraphics[width=0.45\textwidth]{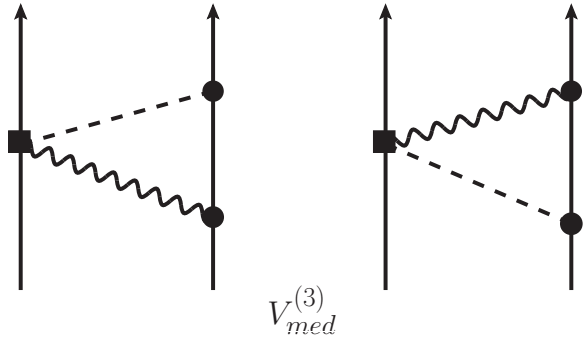}
\caption{Double exchanges generating the contribution $V_\text{med}^{(3)}$ to
the in-medium NN-interaction. Mirror graphs $N_1\leftrightarrow N_2$ have to be
supplemented.}
\end{figure}

In chiral effective field theory, 3N-interactions depend on the spins $\vec \sigma_j$, isospins $\vec \tau_j$, and momenta of the three involved nucleons. Denoting the in-going and out-going nucleon momenta by $\vec p_j$ and $\vec p_j\!\!\,'$, $(j = 1,2,3)$, the associated momentum-transfers  $\vec q_j=\vec p_j\!\!\,'-\vec p_j$ satisfy the relation  $\vec q_1+\vec q_2+\vec q_3=0$. In order to determine the in-medium NN-interaction $V_\text{med}$, we consider the on-shell scattering process $N_1(\vec p\,)+ N_2(-\vec p\,) \to N_1(\vec p\,')+N_2(-\vec p\,')$ in the center-of-mass frame, such that $|\vec p\,|=|\vec p\,'|=p$. The in-going and outgoing momenta are $\pm \vec p$ and $\pm \vec p\,'$, and $\vec q=\vec p\,'-\vec p$ is the momentum transfer with $0\leq q\leq2p$. For NN-scattering in 
isospin-symmetric spin-saturated nuclear matter of density $\rho=2k_f^3/3\pi^3$, the isospin operators appearing in $V_\text{med}$ are the same as in free space, namely $1$ and $\vec\tau_1\!\cdot\! \vec\tau_2$. Although Galilei invariance is broken by the nuclear medium, this similarity applies for the chosen on-shell kinematics ($|\vec p\,|=|\vec p\,'|$) also to the spin-momentum operators.\footnote{Going off-shell ($|\vec p\,|\ne|\vec p\,'|$) introduced some spin-momentum operators that are unfamilar from the NN-potential. In practice the off-shell extrapolation \cite{holt} of $V_\text{med}$ is performed by the substitution $p^2 \to (p^2+p'^2)/2$ (or $pp'$).}   Therefore one can expand $V_\text{med}$ in terms of the following five independent spin operators: 
\begin{equation}  1\,, \quad \vec\sigma_1\!\cdot\!\vec\sigma_2\,, \quad 
\vec\sigma_1\!\cdot\!\vec q\, \vec\sigma_2\!\cdot\! \vec q\,, \quad i(\vec\sigma_1\!+\!\vec\sigma_2)\!\cdot\!(\vec q\!\times\!\vec p\,)\,, \quad 
\vec\sigma_1\!\cdot\! \vec p\, \vec\sigma_2\!\cdot\!\vec p +\vec \sigma_1\!\cdot\! \vec p\,' \vec \sigma_2\!\cdot\!\vec p\,'\,. \end{equation} 
The quadratic spin-orbit operator will be used occasionally for notational convenience and it possesses the following decomposition:
\begin{equation}
\vec\sigma_1\!\cdot\!(\vec q\!\times\!\vec p\,)\vec\sigma_2\!\cdot\!(\vec q\!\times\!\vec p\,)= 
q^2\Big(p^2 -{q^2 \over 4}\Big) \vec\sigma_1\!\cdot\!\vec\sigma_2 +\Big({q^2 \over 2}-p^2\Big)
 \vec\sigma_1\!\cdot\!\vec q\, \vec\sigma_2\!\cdot\! \vec q-{q^2 \over 2}(\vec\sigma_1\!\cdot\! \vec p\, \vec\sigma_2\!\cdot\!\vec p +\vec \sigma_1\!\cdot\! \vec p\,' \vec \sigma_2\!\cdot\!
 \vec p\,')\,. \end{equation}
As a side remark, we note that the sign-convention for the NN-potential is chosen such that 
ordinary $1\pi$-exchange gives $V_{1\pi} = -(g_A/2f_\pi)^2(m_\pi^2+q^2)^{-1}  \vec\tau_1\!\cdot\! \vec\tau_2 \, \vec\sigma_1\!\cdot\!\vec q\, \vec\sigma_2\!\cdot\! \vec q$. The occurring parameters are  $g_A=1.3$, $f_\pi = 92.2\,$MeV and $m_\pi = 138\,$MeV.
After these preparations, we turn now to the enumeration of the contributions to the in-medium NN-potential $V_\text{med}$, following the presentation of subleading chiral 3N-forces $V_\text{3N}$ in ref.\,\cite{3Nshort}.
\section{One-pion-exchange-contact topology}
For the $1\pi$-exchange-contact topology two nonzero contributions to $V_\text{3N}$ have been derived in section\,II of ref.\cite{3Nshort}, which we take from eq.(2.1):
\begin{equation} V_\text{3N} =-{g_A^4C_T m_\pi \over 16\pi f_\pi^{4}}{\vec\tau_1\!
 \cdot\! \vec\tau_2 \over m_\pi^2+q_1^2} \,\vec \sigma_1\!\cdot\!
  \vec q_1\,\vec \sigma_3\!\cdot\!\vec q_1\,, \end{equation}
and  eq.(2.3):
\begin{equation} V_\text{3N}={g_A^4 C_Tm_\pi \over 32 \pi f_\pi^4} {\vec \sigma_1
\cdot \vec q_1\over m_\pi^2+ q_1^2} \big[2\vec\tau_1\!\cdot\!\vec\tau_3\,
\vec \sigma_3\!\cdot\!\vec q_1 -\vec\tau_1\!\cdot\!(\vec\tau_2\!\times\!\vec
\tau_3)\, (\vec\sigma_2\!\times\! \vec\sigma_3)\!\cdot\! \vec q_1\big]\,.
\end{equation}
We remind that the parameter $C_T$ belongs to the leading order NN-contact potential $V_\text{ct}=C_S+C_T\vec\sigma_1\!\cdot\!\vec\sigma_2$ and the factor $m_\pi/8\pi$ in $V_\text{3N}$ stems from  pion-loop integrals evaluated in dimensional regularization. Computing all the in-medium diagrams with self closings, vertex corrections and double exchanges (shown in Figs.\,2,\,3,\,4,\,5) for both 3N-interactions in eqs.(3,4) and summing up the individual contributions to 
$V_\text{med}$, one obtains the following total result:
\begin{eqnarray} 
V_\text{med}&=& (1+2-3){g_A^4 C_Tm_\pi k_f^3 \over 24 \pi^3 f_\pi^4}
{\vec \tau_1\!\cdot\!\vec \tau_2 \over m_\pi^2+q^2}\, \vec\sigma_1\!\cdot\!
\vec q\, \vec\sigma_2\!\cdot\! \vec q \nonumber \\
&& +{g_A^4 C_Tm_\pi \over 48 \pi^3 f_\pi^4}(2k_f^3-3m_\pi^2\Gamma_0)\big[
   \vec\tau_1\!\cdot\!\vec \tau_2(1-2\vec\sigma_1\!\cdot\! \vec
\sigma_2)-3\big] \nonumber \\
&& +{g_A^4 C_Tm_\pi \over 32\pi^3 f_\pi^4}(3+\vec\tau_1\!\cdot\! \vec\tau_2)
\big[(\Gamma_0+2\Gamma_1+\Gamma_3)(\vec\sigma_1\!\cdot\! \vec p\, \vec\sigma_2
\!\cdot\!\vec p +\vec \sigma_1\!\cdot\! \vec p\,'\vec \sigma_2\!\cdot\!
\vec p\,')+2\Gamma_2\,\vec \sigma_1\!\cdot\! \vec\sigma_2\big]\,.
\end{eqnarray} 
The first line is obviously zero. From the known eigenvalues $\vec\sigma_1\!\cdot\! \vec
\sigma_2 \to 4S-3$ and $\vec\tau_1\!\cdot\!\vec \tau_2 \to 4I-3$, with $S=0,1$ the total spin and 
$I=0,1$ the total isospin of the two-nucleon system, one concludes that the second line contributes only in the $^1S_0$ partial wave. The term in the third line is operative only for total isospin $I=1$ and partial wave projection shows furthermore that it contributes only to the $s$-wave. Therefore, the in-medium potential remaining in the $^1S_0$ partial wave is:
\begin{equation}V_\text{med}^{(^1S_0)} = {g_A^4 C_Tm_\pi \over 4\pi^3 f_\pi^4}
\bigg\{ {2k_f^2 \over 3}-m_\pi^2\Gamma_0-p^2( \Gamma_0+2\Gamma_1+
\Gamma_3)-3\Gamma_2\bigg\}=0\,,\end{equation}
and it vanishes by the identity:
\begin{equation}\int\limits_{|\vec l\,|<k_f}\!\!\!\!{d^3 l \over 2\pi} \, {m_\pi^2
+(\vec l+\vec p\,)^2 \over m_\pi^2+(\vec l+ \vec p\,)^2} = {2k_f^3 \over 3} =
m_\pi^2\Gamma_0+3\Gamma_2 +p^2 (\Gamma_3+2\Gamma_1+ \Gamma_0)\,.
\end{equation}
The last part in eq.(7) follows by expanding the numerator and expressing the Fermi sphere integrals in terms of the loop functions $\Gamma_\nu(p,k_f)$, which are given in explicit form in the appendix. The effective vanishing of $V_\text{med}$ from the $1\pi$-exchange-contact topology is an important consistency check, since the underlying $V_\text{3N}$ vanishes after complete antisymmetrization as demonstrated at the end of section\,II in ref.\,\cite{3Nshort}.

\section{Two-pion-exchange-contact topology}
Next, we come to the $2\pi$-exchange-contact topology, for which ref.\cite{3Nshort} has derived two contributions to 3N-interaction $V_\text{3N}$. The structurally simpler one written in 
eq.(3.5) has the form:
\begin{equation}
V_\text{3N} = -{g_A^2 C_T \over 24 \pi f_\pi^4} \vec\tau_1\!\cdot\! \vec\tau_2
\,  \vec\sigma_2\!\cdot\! \vec\sigma_3 \big[m_\pi+(2m_\pi^2+q_1^2)A(q_1)\big]\,,
\end{equation}
with the pion-loop function
\begin{equation} A(q_1)={1\over 2q_1}\arctan{q_1 \over 2m_\pi}\,.\end{equation}
The corresponding self-closing term $V_\text{med}^{(0)}=0$ due to vanishing spin- or isospin-traces and the short-range vertex corrections introduce merely a factor $-3\rho$, leading to the result: 
\begin{equation} V_\text{med}^{(1)} = {g_A^2 C_Tk_f^{3} \over 12 \pi^3 f_\pi^4} 
\vec\tau_1\!\cdot\!\vec\tau_2\big[m_\pi+(2m_\pi^2+q^2)A(q)\big]\,.\end{equation}
Pionic vertex corrections and double exchanges differ only with respect to the isospin operator and therefore can be combined to one single expression:
\begin{eqnarray} 
  V_\text{med}^{(2)}+V_\text{med}^{(3)}&=&{g_A^2 C_T\over 720 \pi^3 f_\pi^4}\,
\vec\sigma_1\!\cdot\!\vec\sigma_2 (3+\vec\tau_1\!\cdot\!\vec\tau_2)
\bigg\{k_f m_\pi(p^2+11k_f^2-4 m_\pi^2)\nonumber\\ &&+(p+k_f)^2\bigg[{k_f\over p}
(k_f^2+10m_\pi^2) -5m_\pi^2+{7k_f^2 \over 4}+{k_f p\over 2}- {p^2 \over 4}\bigg]
 \arctan{p+k_f \over 2m_\pi}\nonumber\\ &&+(p-k_f)^2\bigg[{k_f\over p}
(k_f^2+10m_\pi^2) +5m_\pi^2-{7k_f^2 \over 4}+{k_f p\over 2}+ {p^2 \over 4}\bigg]
\arctan{p-k_f \over 2m_\pi}\nonumber\\ && +{m_\pi^3\over p}(5p^2-5k_f^2+4 m_\pi^2)
\ln{4 m_\pi^2 + (p+k_f)^2 \over 4 m_\pi^2 + (p-k_f)^2}\bigg\}\,.\end{eqnarray} 
Note that this term acts only in the $^1S_0$ partial wave with $\vec\sigma_1\!\cdot\!\vec
\sigma_2 (3+\vec\tau_1\!\cdot\!\vec\tau_2) \to -12$. 

The other 3N-interaction provided by the $2\pi$-exchange-contact topology reads according to 
eq.(3.1) in ref.\,\cite{3Nshort}: 
\begin{eqnarray} 
  V_\text{3N} &=& {g_A^4 C_T \over 48 \pi f_\pi^4} \bigg\{2 \vec\tau_1\!\cdot\!
\vec\tau_2\, \vec\sigma_2\!\cdot\! \vec\sigma_3\bigg[3m_\pi-{m_\pi^3\over 4
m_\pi^2+q_1^2}+2(2m_\pi^2+q_1^2)A(q_1)\bigg] \nonumber \\ && +9\big[\vec\sigma_1
\!\cdot\!\vec q_1\,\vec \sigma_2\!\cdot \!\vec q_1-q_1^2\, \vec\sigma_1\!\cdot\!
 \vec\sigma_2\big] A(q_1)\bigg\}\,,\end{eqnarray}
where the first part is structurally equivalent to eq.(8). It is advantageous to combine
self closings and short-range vertex corrections: 
\begin{eqnarray}V_\text{med}^{(0)} + V_\text{med}^{(1)} &=& {g_A^4 C_Tk_f^3 \over 12 \pi^3 
f_\pi^4} \bigg\{{3 \over 2}(\vec\sigma_1\!\cdot\!\vec q\, \vec\sigma_2\!\cdot\! \vec q
-q^{2}\, \vec\sigma_1\!\cdot\! \vec\sigma_2)A(q)\nonumber \\ && +\vec\tau_1 \!
\cdot\! \vec \tau_2 \bigg[{m_\pi^3 \over 4m_\pi^2+q^2}-3m_\pi -2(2m_\pi^2+q^2)
 A(q)\bigg]\bigg\}\,,\end{eqnarray}
since for these the medium effect amounts to merely a factor of density $\rho=2k_f^3/3\pi^2$. 
The contributions from pionic vertex corrections and double exchanges are also preferably combined into one expression, which reads: 
\begin{eqnarray}   V_\text{med}^{(2)}+V_\text{med}^{(3)}&=& {g_A^4 C_T\over
40 \pi^3 f_\pi^4}\bigg[{H_1\over 2}- {H_0\over 9}(3+\vec\tau_1\!\cdot\!
\vec \tau_2)\vec\sigma_1\!\cdot\! \vec\sigma_2 \bigg]\nonumber \\ &&
  + {g_A^4 C_T \over 64 \pi^3 f_\pi^4} \Big[H_2\,\vec\sigma_1\!\cdot\!
\vec\sigma_2+H_3 (\vec\sigma_1\!\cdot\! \vec p\, \vec\sigma_2\!\cdot\!
\vec p +\vec\sigma_1\!\cdot\! \vec p\,' \vec\sigma_2\!\cdot\!\vec p\,')\Big]\,.
\end{eqnarray}
The occurring $(p,k_f)$-dependent functions $H_{0,1,2,3}$ have the following analytical forms:
\begin{eqnarray}  H_0&=& k_f m_\pi(p^2+21k_f^2-19 m_\pi^2)+{m_\pi^3\over
 4p}(35p^2-35k_f^2-44 m_\pi^2)\ln{4 m_\pi^2+(p+k_f)^2\over 4 m_\pi^2 + (p-k_f)^2}
\nonumber\\ &&+\bigg[{k_f^3\over p}(k_f^2+10m_\pi^2) -5p^2 m_\pi^2+5 k_f^3 p
+{5\over 2}p^2 k_f^2\nonumber\\ &&+15m_\pi^2(k_f^2+2m_\pi^2)+{1\over 4}(15k_f^4
-p^4)\bigg] \arctan{p+k_f \over 2m_\pi}\nonumber\\ &&+\bigg[{k_f^3\over p}
(k_f^2+10m_\pi^2)+5p^2 m_\pi^2+5 k_f^3 p -{5\over 2}p^2 k_f^2\nonumber\\ &&
-15m_\pi^2 (k_f^2+2m_\pi^2)+{1\over 4}(p^4-15k_f^4)\bigg] \arctan{p-k_f \over
 2m_\pi}\,,\end{eqnarray}

\begin{eqnarray}  H_1&=& k_f m_\pi(p^2-9k_f^2+36 m_\pi^2)+{2m_\pi^3\over p}
(5k_f^2+12 m_\pi^2-5p^2)\ln{4 m_\pi^2+(p+k_f)^2\over 4 m_\pi^2 + (p-k_f)^2}
\nonumber\\ &&+\bigg[{k_f^5\over p}-60m_\pi^4 +5 k_f^3 p+{5\over 2}p^2 k_f^2
+{1\over 4}(15k_f^4-p^4)\bigg] \arctan{p+k_f \over 2m_\pi}\nonumber\\ &&
+\bigg[{k_f^5\over p}+60m_\pi^4 +5 k_f^3 p-{5\over 2}p^2 k_f^2
+{1\over 4}(p^4-15k_f^4)\bigg] \arctan{p-k_f \over 2m_\pi}\,,\end{eqnarray} 

\begin{eqnarray}  H_2&=&{k_f m_\pi \over 7p^2}\bigg[{19 k_f^4\over 10}
 +{16 k_f^2\over 15}(6m_\pi^2-19p^2)+8m_\pi^4 +68p^2m_\pi^2+{p^4 \over 2}\bigg]
+ { m_\pi \over 8p^3}\bigg[ -{64\over 7} m_\pi^6\nonumber\\ && + {48\over 5}
m_\pi^4(5p^2-k_f^2) +4m_\pi^2(6k_f^2p^2- 5p^4-k_f^4)+(p^2-k_f^2)^3
\bigg] \ln{4 m_\pi^2+(p+k_f)^2\over 4 m_\pi^2 + (p-k_f)^2}
\nonumber\\ &&+\bigg[{2k_f^7\over 35p^3}+2 k_f^3 p+{6\over 5}p^2 k_f^2
+k_f^4-{p^4\over 7}-16 m_\pi^4\bigg] \arctan{p+k_f \over 2m_\pi}\nonumber\\ &&
+\bigg[{2k_f^7\over 35p^3}+2 k_f^3 p-{6\over 5}p^2 k_f^2-k_f^4+{p^4\over 7}
+16 m_\pi^4\bigg] \arctan{p-k_f \over 2m_\pi}\,,\end{eqnarray}

\begin{eqnarray}  H_3&=&{k_f m_\pi \over 140p^4}\big[41 p^4+8p^2(13k_f^2-3m_\pi^2)
-3 (19 k_f^4 + 64 k_f^2 m_\pi^2+80 m_\pi^4)\big]\nonumber\\ &&+ {m_\pi\over p^5}
\bigg[{m_\pi^2\over 4}(3k_f^4-2k_f^2p^2-p^4)+{3\over 5} m_\pi^4 (3k_f^2+p^2)+
{12\over7}m_\pi^6+{3\over 16}(k_f^2-p^2)^3\bigg] \nonumber\\ && \times \ln
{4m_\pi^2+(p+k_f)^2\over 4 m_\pi^2+(p-k_f)^2}+ {(p+k_f)^4\over 35 p^5} \big(4 p^3
-16 k_f p^2 +12k_f^2p-3k_f^3\big)\arctan{p+k_f \over 2m_\pi}\nonumber\\ &&
- {(p-k_f)^4\over 35 p^5} \big(4 p^3+16 k_fp^2+12k_f^2p+3 k_f^3\big)\arctan{p-k_f
\over 2m_\pi} \,.\end{eqnarray}
These results have been obtained by using the following reduction formulas for Fermi sphere integrals over even functions $F(s)=F(-s)$: 
\begin{equation} \int\limits_{|\vec l\,|<k_f}\!\!\!\!{d^3 l \over 2\pi} \,
F(|\vec l +\vec p\,|)\big\{1,l_i,l_il_j\big\}=\int_{p-k_f}^{p+k_f}\!\!ds\, {s\over
2p}\big[k_f^2-(p-s)^2 \big]F(s)\big\{1, \chi_1\, p_i,\chi_2\,\delta_{ij}+\chi_3
\, p_ip_j\big\}\,, \end{equation}
with the polynomial weighting-functions:
\begin{eqnarray} &&\chi_1 ={1\over 4p^2}(s^2+2s p-3p^2-k_f^2)\,, \\
&& \chi_2 ={1\over 24p^2}\big[k_f^2-(p-s)^2 \big](s^2+4s p+p^2-k_f^2)\,, \\
&& \chi_3 ={1\over 8p^4}\big[k_f^4+2k_f^2(p^2-s p-s^2)+(p-s)^2(s^2+4s p+5p^2)\big]
\,. \end{eqnarray} 
Note that the analytical result in eq.(11) was obtained with the first of the three reduction formulas.
\section{Leading relativistic corrections}
Next, we treat the relativistic $1/M$-corrections to the chiral 3N-interaction, which can be subdivided into diagrams of $1\pi$-exchange-contact topology (with parameter combination $g_A^2 C_{S,T}/f_\pi^2$) and  diagrams of $2\pi$-exchange topology (proportional to $g_A^{2,4}/
f_\pi^4$). The corresponding expressions for $ V_\text{3N}$ depend on constants $\bar 
\beta_{8,9}$ which parametrize a unitary ambiguity of these 3N-potentials. In order to be consistent with the underlying NN-potential, we follow ref.\,\cite{n4lo2} and choose the values $\bar \beta_8 =1/4$ and $\bar \beta_9 =-1/4$. One should note here that the misprint 
$\bar \beta_9 =0$ in eq.(4.14) of ref.\cite{3Nshort} has been corrected in ref.\,\cite{n4lo2}.
\subsection{$1\pi$-exchange-contact topology}
The $1/M$-correction to the $\pi$N-coupling combined with the 4N-contact vertex $(\sim C_{S,T}$) leads to the 3N-interaction written in eq.(4.6) of ref.\,\cite{3Nshort}, and setting $\bar \beta_9 =-1/4$ it reads: 
\begin{eqnarray}
 V_\text{3N}&=&-{g_A^2\over 16M f_\pi^2}{\vec\tau_1\!\cdot\! \vec\tau_2\over
m_\pi^2+q_1^2}\Big\{C_T\Big[i \vec\sigma_1\!\cdot\!(\vec p_1+\vec p_1\,\!\!')\,
(\vec  \sigma_2\!\times\!\vec \sigma_3)\!\cdot\! \vec q_1+3\vec\sigma_1\!\cdot\!
\vec q_1\,\vec\sigma_3\!\cdot\! \vec q_3\nonumber\\ &&\qquad \qquad \qquad
   \qquad +3i\vec \sigma_1
\!\cdot\!\vec q_1\,(\vec \sigma_2\!\times\!\vec\sigma_3)\!\cdot\!(\vec p_2+\vec
p_2\,\!\!') \Big] +3C_S \,\vec\sigma_1\!\cdot\! \vec q_1\, \vec\sigma_2\!\cdot\!
\vec q_3\Big\}\,.\end{eqnarray}
While self closings vanish (tr\,$\vec \tau_{1,2}=0$, tr\,$\vec \sigma_3=0$ or $\vec q_3=0$), the 
short-range vertex corrections pro- duce a $1\pi$-exchange NN-potential modified by a factor linear in density $\rho$: 
 \begin{equation} V_\text{med}^{(1)}={g_A^2k_f^3 \over 16\pi^2M f_\pi^2}(C_T-C_S)
  {\vec\tau_1\!\cdot\!\vec \tau_2 \over m_\pi^2+q^2}\,\vec\sigma_1\! \cdot\!\vec
  q \, \vec\sigma_2\!\cdot\!\vec q\,.\end{equation}
Furthermore, one gets from pionic vertex corrections the contribution:
\begin{eqnarray}
V_\text{med}^{(2)}&=&{3g_A^2 C_T\over 8\pi^2 M f_\pi^2}\Big\{\Big[2p^2(\Gamma_3-\Gamma_0)
+{3q^2\over 4}(\Gamma_0+ \Gamma_1)+4\Gamma_2\Big]\vec\sigma_1\!\cdot\! \vec\sigma_2 \nonumber\\ &&
-{3\over 2}(\Gamma_0+\Gamma_1)\,\vec\sigma_1\!\cdot\! \vec q \,\vec\sigma_2\!\cdot\! \vec q+
(\Gamma_0-\Gamma_3)(\vec\sigma_1\!\cdot\! \vec p\, \vec\sigma_2 \!\cdot\!\vec p +\vec \sigma_1\!
\cdot\! \vec p\,' \vec\sigma_2\!\cdot\!\vec p\,')\Big\} \nonumber\\ &&+ {9g_A^2 \over 32\pi^2
M f_\pi^2}(\Gamma_0+ \Gamma_1) \Big[(C_S-C_T)i(\vec\sigma_1\!+\!\vec\sigma_2)\!\cdot\!(\vec q\!\times\!\vec p\,)-C_S\,q^2\Big]\,,\end{eqnarray}
and from double exchanges the contribution:
\begin{eqnarray}
V_\text{med}^{(3)}&=&{g_A^2\vec\tau_1\!\cdot\!\vec\tau_2\over 16\pi^2M f_\pi^2}\vec
\sigma_1\!\cdot\!\vec\sigma_2\Big\{C_T\Big[{3q^2\over 2}(\Gamma_0+\Gamma_1)-p^2(5\Gamma_0
 + 6\Gamma_1 +\Gamma_3)\Big]-(2C_T+3C_S)\Gamma_2\Big\}\nonumber\\ && +{g_A^2\vec\tau_1\!
\cdot\!\vec\tau_2\over 32\pi^2M f_\pi^2}\Big\{3(C_S-C_T)(\Gamma_0+ \Gamma_1)\vec\sigma_1 
\!\cdot\!\vec q\, \vec\sigma_2\!\cdot\! 
\vec q \nonumber\\ && + \Big[C_T(5\Gamma_0+6\Gamma_1+\Gamma_3)-3C_S(\Gamma_0+2\Gamma_1
 +\Gamma_3)\Big] (\vec\sigma_1\! \cdot\!\vec p\, \vec\sigma_2\!\cdot\!\vec p +
 \vec\sigma_1\!\cdot\!\vec p\,' \vec\sigma_2\!\cdot\!\vec p\,')\Big\}\nonumber\\ && 
 +{3g_A^2\vec\tau_1\!\cdot\!\vec\tau_2 \over 16\pi^2M f_\pi^2}C_T 
\Big\{(\Gamma_0+\Gamma_1)\Big[{q^2 \over 2} -i(\vec\sigma_1 \!+\!\vec\sigma_2)\!\cdot\!
 (\vec q\!\times\!\vec p\,)\Big]-3\Gamma_2 -p^2(\Gamma_0+2\Gamma_1+\Gamma_3)\Big\}\,, \end{eqnarray}
 which both include a spin-orbit term $\sim i(\vec\sigma_1 \!+\!\vec\sigma_2)\!\cdot\!
 (\vec q\!\times\!\vec p\,)$.

The retardation correction to the $1\pi$-exchange-contact diagram leads to the 3N-interaction written in eq.(4.3) of ref.\,\cite{3Nshort}, and setting  $\bar \beta_8=1/4$ it reads:
\begin{eqnarray}
V_\text{3N}&=&{g_A^2\over 16M f_\pi^2}{\vec\sigma_1\!\cdot\!\vec q_1\, \vec\tau_1\!\cdot\! 
\vec\tau_2\over (m_\pi^2+q_1^2)^2}\Big\{\vec q_1\!\cdot\!\vec q_3(C_S\,\vec\sigma_2\!\cdot\!
\vec q_1+C_T\,\vec\sigma_3\!\cdot\!\vec q_1)\nonumber\\ &&+i C_T(\vec\sigma_2\!\times\! 
\vec\sigma_3)\!\cdot\!\vec q_1\,(3\vec p_1+ 3\vec p_1\,\!'+\vec p_2+\vec p_2\,\!')\!\cdot 
\!\vec q_1\Big\}\,. \end{eqnarray}
Again, $V_\text{med}^{(0)} =0$ and for the contributions from vertex corrections and double exchanges one finds the results: 
\begin{equation}
V_\text{med}^{(1)}= {g_A^2 k_f^3 \over 48 \pi^2 M f_\pi^2}(C_S-C_T)q^2{\vec\tau_1\!\cdot\! \vec\tau_2\over (m_\pi^2+q^2)^2}\,\vec \sigma_1\!\cdot\! \vec q\, \vec\sigma_2\!\cdot\! \vec q \,,
\end{equation}

\begin{eqnarray}V_\text{med}^{(2)}& =& {3g_A^2 C_S\,q^2 \over 32 \pi^2 M f_\pi^2} \big[
\Gamma_0+ \Gamma_1-m_\pi^2(\gamma_0+ \gamma_1)\big] \nonumber\\ &&+ {3g_A^2 C_T\over 8 \pi^2 M f_\pi^2}\Big\{(\gamma_2+\gamma_4)\vec \sigma_1\!\cdot\! \vec q\,\vec\sigma_2\!\cdot\! \vec q +
\Big[2\Gamma_2-{4k_f^3\over 3}+2m_\pi^2(2\Gamma_0-m_\pi^2\gamma_0-\gamma_2)\nonumber\\ &&+
{q^2\over 4}(\gamma_2+\gamma_4) +\Big({q^2\over 4}-4p^2\Big)(m_\pi^2\gamma_0+m_\pi^2\gamma_1
+\gamma_2+\gamma_4-\Gamma_0- \Gamma_1)\Big] \vec\sigma_1\!\cdot\!\vec\sigma_2\nonumber\\ &&+
\Big[\Gamma_0+ 2\Gamma_1+\Gamma_3-m_\pi^2(\gamma_0+ 2\gamma_1+\gamma_3)-4(\gamma_2+\gamma_4) \nonumber\\ && +\Big({q^2\over 8}-2p^2\Big)(\gamma_0+ 3\gamma_1+3\gamma_3+\gamma_5)\Big]
(\vec\sigma_1\!\cdot\!\vec p\, \vec\sigma_2\!\cdot\!\vec p +\vec\sigma_1\!\cdot\!\vec p\,' 
\vec\sigma_2\!\cdot\!\vec p\,')\Big\}\,,\end{eqnarray}

\begin{eqnarray}V_\text{med}^{(3)}& =& {g_A^2 \vec\tau_1\!\cdot\! \vec\tau_2 \over 16 \pi^2 
M f_\pi^2}(C_T-C_S)(\gamma_2+\gamma_4)\vec \sigma_1\!\cdot\! \vec q\,\vec\sigma_2\!\cdot\!\vec q 
\nonumber\\ &&+{g_A^2 C_S \over 32 \pi^2 M f_\pi^2}\vec\tau_1\!\cdot\! \vec\tau_2 \Big\{ \big[ 2\Gamma_2-(2m_\pi^2+q^2)\gamma_2-q^2\gamma_4\big] \vec\sigma_1\!\cdot\!\vec\sigma_2+ \Big[ \Gamma_0+ 2\Gamma_1+\Gamma_3 \nonumber\\ &&-m_\pi^2(\gamma_0+ 2\gamma_1+\gamma_3)-{q^2\over 2}(\gamma_0+3\gamma_1+3\gamma_3+\gamma_5)\Big](\vec\sigma_1\!\cdot\!\vec p\, \vec\sigma_2\!
\cdot\!\vec p +\vec\sigma_1\!\cdot\!\vec p\,' \vec\sigma_2\!\cdot\!\vec p\,')\Big\}\nonumber
\\ &&+{g_A^2C_T\over 16 \pi^2M f_\pi^2}\vec\tau_1\!\cdot\! \vec\tau_2 \Big\{{2k_f^3\over 3}
+m_\pi^2(m_\pi^2\gamma_0- 2\Gamma_0)+{q^2\over 2}\big[m_\pi^2(\gamma_0+ \gamma_1)-\Gamma_1
\big]\nonumber \\ &&+ \Big[\Big(4p^2+{q^2\over 2}\Big)\big(m_\pi^2(\gamma_0+\gamma_1)+\gamma_2
+\gamma_4-\Gamma_0-\Gamma_1\big)\nonumber \\ && +2k_f^3-3\Gamma_2+3 m_\pi^2(\gamma_2+m_\pi^2\gamma_0-2\Gamma_0)\Big] \vec\sigma_1\!\cdot\!\vec\sigma_2 \nonumber \\ &&+
\Big[{3\over 2}\Big(m_\pi^2(\gamma_0+ 2\gamma_1+\gamma_3)-\Gamma_0- 2\Gamma_1-\Gamma_3\Big)+ 4(\gamma_2+\gamma_4)\nonumber \\ && +\Big(2p^2+{q^2\over 4}\Big)(\gamma_0+ 3\gamma_1+3\gamma_3+\gamma_5)\Big](\vec\sigma_1\!\cdot\!\vec p\, \vec\sigma_2\!\cdot\!\vec p +\vec\sigma_1\!\cdot\!\vec p\,' \vec\sigma_2\!\cdot\!\vec p\,')\Big\}
\,.\end{eqnarray}
The new loop functions $\gamma_\nu(p,k_f)$ are defined by Fermi sphere integrals over a squared pion propagator and their analytical expressions (involving arctangents and logarithms) are given in the appendix. 
\subsection{$2\pi$-exchange topology}
The $1/M$-correction to the isovector Weinberg-Tomozawa $\pi\pi$NN-vertex combined with two ordinary $\pi$N-couplings gives rise to the 3N-interaction written in eq.(4.8) of ref.\,\cite{3Nshort}: 
\begin{equation} V_\text{3N}=-{g_A^2 \over 16M f_\pi^4}{\vec\sigma_1\!\cdot\! \vec q_1 \,\vec\sigma_3\!\cdot\! \vec q_3\over (m_\pi^2+q_1^2)(m_\pi^2+q_3^2)}\,\vec\tau_1\!\cdot\!(\vec\tau_2 
\!\times\! \vec\tau_3)\Big[\vec \sigma_2\!\cdot\!(\vec q_{1}\!\times\!\vec q_3)+{i \over 2}
(\vec p_ 2 +\vec p_2\,\!\!')\!\cdot\!(\vec q_3-\vec q_1)\Big]\,.
\end{equation}
Here, we have multiplied $V_\text{3N}$ with a factor 2, since it is symmetric under 
$1\leftrightarrow 3$ to account already for this permutation. Alternatively, one could  distinguish the equivalent couplings to nucleon\,1 and nucleon\,3 and the factor 2 would emerge at the end of the calculation. Obviously, the self closings vanish (tr\,$\vec \tau_{1,2,3}=0$), while one gets the following contributions from pionic vertex corrections:
 
\begin{equation}
V_\text{med}^{(1)}={g_A^2 \over 32 \pi^2M f_\pi^4}\, {\vec\tau_1\!\cdot\!\vec\tau_2  \over 
m_\pi^2+q^2} \Big[13\Gamma_2-m_\pi^2 \Gamma_1+3p^2(\Gamma_0+ 2\Gamma_1+\Gamma_3)\Big]
\vec\sigma_1 \!\cdot\! \vec q \,\vec\sigma_2\!\cdot\! \vec q \,, \end{equation}
and double exchanges:
\begin{eqnarray}
V_\text{med}^{(3)}&=& {g_A^2\vec\tau_1\!\cdot\! \vec\tau_2 \over 16 \pi^2M f_\pi^4}\Big\{ \Big(p^2-{q^2\over 4}\Big)\big[(2m_\pi^2+q^2)(G_0+2G_1)-2\Gamma_0-2\Gamma_1\big]\nonumber \\ 
&& +G_2 (q^2 \vec \sigma_1\!\cdot\!\vec\sigma_2-\vec\sigma_1\!\cdot\! \vec q\, \vec\sigma_2\!\cdot\!\vec q\,) +(G_0+4G_1+4G_3)\,\vec\sigma_1\!\cdot\!(\vec q\!\times\!\vec p\,)\vec\sigma_2 
\!\cdot\!(\vec q\!\times\!\vec p\,)\nonumber \\ &&+ \Big[\Big({m_\pi^2+q^2 \over 2}-p^2\Big)
G_0+\Big(m_\pi^2+{3q^2\over 2}-4p^2\Big)G_1-G_2\nonumber \\ &&+(q^2-4p^2)G_3 
-{1\over 2}(\Gamma_0+ \Gamma_1)\Big]i(\vec\sigma_1\!+\!\vec\sigma_2)\!\cdot\!(\vec q 
\!\times\!\vec p\,)\Big\}\,.  \end{eqnarray}
The loop functions $G_\nu(p,q,k_f)$ are defined by Fermi sphere integrals over the product of two different pion propagators and given in the appendix.

The retardation correction to the above $2\pi$-exchange mechanism leads to the 3N-interaction written in eq.(4.2) of ref.\,\cite{3Nshort}: 
\begin{equation}
V_\text{3N}={ig_A^2\over 32M f_\pi^4}{\vec\sigma_1\!\cdot\! \vec q_1\,\vec\sigma_3\!\cdot\! \vec q_3 \over (m_\pi^2+q_1^2)(m_\pi^2+q_3^2)}\vec\tau_1\!\cdot\!(\vec\tau_2\!\times\!\vec\tau_3) \big[\vec q_3\!\cdot\!(\vec p_3+\vec p_3\!\,')-\vec q_1\!\cdot\!(\vec p_1+\vec p_1\!\,')\big]\,,
\end{equation}
where we have again multiplied $V_\text{3N}$ with a factor 2, due to its symmetry under $1 \leftrightarrow 3$. One finds the following contributions from pionic vertex corrections: 
\begin{equation}
V_\text{med}^{(1)}={g_A^2\over 32 \pi^2M f_\pi^4}\, {\vec\tau_1\!\cdot\! \vec\tau_2\over 
m_\pi^2+ q^2} \Big[m_\pi^2\Gamma_1-\Gamma_2+p^2(\Gamma_0+ 2\Gamma_1+\Gamma_3)\Big] 
\vec\sigma_1\!\cdot\! \vec q\,\vec\sigma_2\!\cdot\! \vec q\,, \end{equation}
and double exchanges:
\begin{eqnarray}
V_\text{med}^{(3)}&=& {g_A^2 \vec\tau_1\!\cdot\! \vec\tau_2\over 32 \pi^2 M f_\pi^4}
\Big\{6\Gamma_2+2p^2(\Gamma_1+\Gamma_3)+(2m_\pi^2+q^2)(p^2 G_0-G_{0*})
\nonumber \\ &&+ \big[G_{0*}-G_{1*}+p^2(G_1-G_0)\big]i(\vec\sigma_1\!+\!\vec\sigma_2)\!\cdot\! (\vec q \!\times\!\vec p\,)\Big\}\,.  \end{eqnarray}
The star in the index of a loop function $G_{\nu*}$  symbolizes an extra factor $\vec l^{\,2}$ in the respective Fermi sphere integral $(2\pi)^{-1}\!\int\!d^3l\,\theta(k_f-|\vec l\,|)$.

The $1/M$-correction to the $\pi$N-coupling leads via the mechanism of two consecutive pion-exchanges to the 3N-interaction written in eq.(4.5) of ref.\,\cite{3Nshort}, and setting $\bar \beta_9=-1/4$ it reads:
\begin{eqnarray}
V_\text{3N}&=&{g_A^4 \over 64M f_\pi^4} {\vec\sigma_1\!\cdot\! \vec q_1 \over (m_\pi^2+q_1^2) 
(m_\pi^2+q_3^2)} \Big\{\vec \tau_1\!\cdot\! \vec\tau_3\big[3\vec\sigma_3\!\cdot\! \vec q_3\big(
               i\vec\sigma_2\!\cdot\!(\vec q_1\!\times\!(\vec p_2+\vec p_2\!\,'))+q_1^2\big)\\ &&+i\vec \sigma_3\!\cdot\!(\vec p_3+\vec p_3\!\,')\vec \sigma_2\!\cdot\!(\vec q_1\!\times\!\vec q_3)\big]-i\vec\tau_1\!\cdot\!(\vec\tau_2\!\times\! \vec\tau_3)\big[3\vec\sigma_3\!\cdot\! \vec q_3 \, \vec q_1\!\cdot\!(\vec p_2+\vec p_2\!\,')+\vec \sigma_3\!\cdot\!(\vec p_3+\vec p_3\!\,')\,\vec q_1 \!\cdot\!\vec q_3\big]\Big\}\,. \nonumber \end{eqnarray}
The contribution from self closings
\begin{equation}
V_\text{med}^{(0)}=-{g_A^4k_f^3\,q^2\over 16 \pi^2 M f_\pi^4} {\vec\tau_1\!\cdot\! 
\vec\tau_2 \over (m_\pi^2+q^2)^2}\vec\sigma_1\!\cdot\! \vec q\, \vec\sigma_2\!\cdot\! 
\vec q\,,\end{equation}
comes exclusively from the $q_1^2$-term in the first line of eq.(37). Moreover, one obtains the 
following contributions from pionic vertex corrections:
\begin{eqnarray}
V_\text{med}^{(1)}&=&{g_A^4\over 128 \pi^2 M f_\pi^4} {\vec\tau_1\!\cdot\! \vec\tau_2\over 
m_\pi^2+q^2}\Big\{\Big[p^2(11\Gamma_0+6\Gamma_1-5\Gamma_3)- q^2(2\Gamma_0+5\Gamma_1+3 \Gamma_3)
\nonumber \\ &&+ 3m_\pi^2\Gamma_1-41\Gamma_2\Big]\vec \sigma_1\!\cdot\! \vec q\, \vec\sigma_2 \!\cdot\! \vec q -q^2(\Gamma_0+2\Gamma_1+ \Gamma_3) (\vec\sigma_1\!\cdot\!\vec p\, \vec\sigma_2\!\cdot\!\vec p +\vec\sigma_1\!\cdot\!\vec p\,' \vec\sigma_2\!
\cdot\!\vec p\,') \Big\}\,, \end{eqnarray}
and double exchanges:
\begin{eqnarray}
V_\text{med}^{(3)}&=&{3g_A^4\over 128 \pi^2 M f_\pi^4} \Big\{4k_f^3-3q^2(\Gamma_0+\Gamma_1)
+3m_\pi^2\big[(2m_\pi^2 +q^2)G_0-4\Gamma_0\big] \nonumber \\ && +\Big[ 2G_2-\Gamma_0+
2\Gamma_1 +(m_\pi^2+2p^2+q^2)G_0+2(4p^2-q^2)(G_1+G_3)\nonumber \\ && -4m_\pi^2G_1\Big] 
i(\vec\sigma_1\! +\! \vec\sigma_2)\!\cdot\!(\vec q\!\times\!\vec p\,)+4(G_0+2G_1)\,
\vec\sigma_1\!\cdot\!(\vec q\!\times\!\vec p\,)\vec\sigma_2 \!\cdot\!(\vec q\!\times\!
\vec p\,)\Big\} \nonumber \\ &&
+{g_A^4 \vec\tau_1\!\cdot\!\vec\tau_2 \over 64 \pi^2 M f_\pi^4} \Big\{(8p^2-3q^2)
(\Gamma_0+\Gamma_1) +{4\over 3}k_f^3-4m_\pi^2\Gamma_0\nonumber \\ && + (2m_\pi^2 +q^2)
\big[ (q^2-4p^2)(G_0+2G_1)+m_\pi^2G_0\big]  
+\Big[ 6G_2-\Gamma_0\nonumber \\ && +(m_\pi^2+6p^2-q^2)G_0+6(4p^2-q^2)(G_1+G_3)\Big] 
i(\vec\sigma_1\! +\! \vec\sigma_2)\!\cdot\!(\vec q\!\times\!\vec p\,)\Big\}\,. \end{eqnarray}
Note that the isoscalar part of $V_\text{med}^{(3)}$ includes a quadratic spin-orbit term $\sim \vec\sigma_1\!\cdot\!(\vec q\!\times\!\vec p\,)\vec\sigma_2 \!\cdot\!(\vec q\!\times\!
\vec p\,)$. For the choice $\bar \beta_9=0$, this term would be absent and furthermore 
$V_\text{med}^{(3)}$ would carry the overall isospin factor $3+2\vec\tau_1\!\cdot\!\vec\tau_2$.  

Finally, there are the retardation corrections to the (consecutive) $2\pi$-exchange. These generate the 3N-interaction written in eq.(4.1) of ref.\,\cite{3Nshort}, and setting $\bar \beta_8=1/4$ it reads: 
\begin{eqnarray}
V_\text{3N}&=&{g_A^4 \over 64M f_\pi^4}{\vec\sigma_1\!\cdot\! \vec q_1\,\vec\sigma_3\!\cdot \! \vec q_3 \over (m_\pi^2+q_1^2)^2(m_\pi^2+q_3^2)}\Big\{-\vec q_1\!\cdot\!\vec q_3\big[\vec\tau_1  \!\cdot\! \vec\tau_3\,\vec q_1\!\cdot\!\vec q_3+\vec\tau_1\!\cdot\!(\vec\tau_2\!\times\!  \vec 
\tau_3)\, \vec\sigma_2\!\cdot\!(\vec q_1\!\times\! \vec q_3)\big] \nonumber \\ && +i\big[ \vec \tau_1 \!\cdot\! \vec\tau_3\, \vec\sigma_2\!\cdot\!(\vec q_1\!\times\! \vec q_3) -\vec\tau_1\!\cdot\!(\vec\tau_2\!\times\! \vec\tau_3)\, \vec q_1 \! \cdot\!\vec q_3 \big]\, \vec q_1\! 
\cdot\!(3\vec p_1 +3\vec p_1\!\,'+\vec p_2+\vec p_2\!\,')\Big\}\,.\end{eqnarray}
The contribution from self closings:
\begin{equation}
V_\text{med}^{(0)}={g_A^4 k_f^3\, q^4 \over 48 \pi^2 M f_\pi^4} {\vec\tau_1\!\cdot\! \vec\tau_2 \over (m_\pi^2+q^2)^3}\, \vec\sigma_1\!\cdot\! \vec q\, \vec\sigma_2\!\cdot\! \vec q\,,\end{equation} 
stems from the first term $\sim (\vec q_1\!\cdot\!\vec q_3)^2$ in eq.(41). The full result from the pionic vertex corrections takes the form: 
\begin{eqnarray}
V_\text{med}^{(1)}&=& {g_A^4 \vec\tau_1\!\cdot\! \vec\tau_2 \over 64 \pi^2 M f_\pi^4}{
\vec\sigma_1\!\cdot\! \vec q\, \vec\sigma_2\!\cdot\! \vec q \over m_\pi^2+q^2} \bigg\{
{q^2 \over m_\pi^2+q^2} \bigg[ 3 (\Gamma_2 + \Gamma_4) - {k_f^3 \over 3} + {q^2 \over 4} 
(\Gamma_0 + 3 \Gamma_1 + 3 \Gamma_3 + \Gamma_5)\bigg] +{8k_f^3 \over 3} \nonumber \\ && + 4\Gamma_2 +4m_\pi^2( m_\pi^2\gamma_0-\gamma_2-2\Gamma_0)+q^2\Big[\Gamma_3+ {\Gamma_1-
\Gamma_0 \over 2} - 2\gamma_2-2\gamma_4+{m_\pi^2 \over 2}( \gamma_0-\gamma_1-2\gamma_3)\Big]
\nonumber \\ && +( 8p^2-q^2) \Big[ m_\pi^2(\gamma_0+\gamma_1)-\Gamma_0 - \Gamma_1 -\gamma_2-\gamma_4 -{q^2 \over 4}( \gamma_0+3\gamma_1+3\gamma_3+\gamma_5) \Big]\bigg\}
\,.\end{eqnarray}
For reasons of clarity, we split the contribution $V_\text{med}^{(3)}$ from double exchanges  into two pieces. Evaluating the 3N-interaction terms in the first line of eq.(41) one obtains: 
\begin{eqnarray}
V_\text{med}^{(3)}&=&{g_A^4 \over 128 \pi^2M f_\pi^4}\bigg\{ {3q^2\over 4}(6\Gamma_1-4 \gamma_2-q^2 \gamma_3)-4k_f^3+{9\over 4}(2m_\pi^2+q^2)(4\Gamma_0-q^2\gamma_1)\nonumber \\ && -{9\over 4}(2m_\pi^2+q^2)^2 (\gamma_0+G_0)+{3\over 8}(2m_\pi^2+q^2)^3K_0 +(2\vec\tau_1\!\cdot\! \vec\tau_2
-3 )\Big[\Gamma_0+\Gamma_1-{q^2\over 4}(\gamma_1+\gamma_3)\nonumber \\ && -\Big(m_\pi^2+{q^2 \over 2}\Big)(\gamma_0+\gamma_1+G_0+2G_1)+{1\over 8}(2m_\pi^2+q^2)^2(K_0+2K_1)\Big] i(\vec\sigma_1\!+\!\vec\sigma_2)\!\cdot\!(\vec q\!\times\!\vec p\,)\nonumber \\ && 
+\vec\tau_1\!\cdot\! \vec\tau_2\Big[(2m_\pi^2+q^2)K_2-2\gamma_2-2G_2\Big](q^2\vec\sigma_1\!\cdot\!\vec\sigma_2-\vec\sigma_1\!\cdot\! \vec q\, \vec\sigma_2\!\cdot\! \vec q\,)
+\vec\tau_1\!\cdot\! \vec\tau_2\Big[(2m_\pi^2+q^2)\nonumber \\ && \times (K_0 +4K_1 +4K_3 ) \ -2(\gamma_0+2\gamma_1+\gamma_3+G_0+4G_1+4G_3)\Big]\vec\sigma_1\!\cdot\!(\vec q\!\times\!\vec p\,)\vec\sigma_2\!\cdot\!(\vec q\!\times\!\vec p\,) \bigg\} \,,\nonumber \\ \end{eqnarray}
and the terms in second line of eq.(41) yield the additional contribution:
\begin{eqnarray}V_\text{med}^{(3)}&=&{g_A^4 \over 128 \pi^2M f_\pi^4}\bigg\{ \vec\tau_1\!\cdot
\! \vec\tau_2 \bigg[q^2\Big(3\Gamma_1-2\gamma_2-{q^2\over 2}\gamma_3 +10\gamma_4+2p^2\gamma_5
\Big) -{8k_f^3 \over 3}-8(3\Gamma_2+p^2\Gamma_3)\nonumber \\ &&+ \Big(3m_\pi^{2}+{3q^2\over 2}+2p^2\Big)(4\Gamma_0-q^2\gamma_1) -(2m_\pi^2+q^2)\Big(3m_\pi^{2}+{3q^2 \over 2 } +4p^2\Big) (\gamma_0+G_0)\nonumber \\ &&+ 4(2m_\pi^2+q^2)(3\gamma_2+p^2\gamma_3+G_{0*})+ (2m_\pi^2+q^2)^2
\Big((2m_\pi^2+q^2+4p^2){K_0\over 4}-K_{0*}\Big) \bigg] \nonumber \\ && +\Big(\vec\tau_1\!
\cdot\! \vec\tau_2 -{3\over 2}\Big)\bigg[{q^2\over 2}(\gamma_1+\gamma_3) +(2m_\pi^2+2p^2+q^2)(\gamma_0+\gamma_1+G_0+2G_1)\nonumber \\ && -2(\Gamma_0+\Gamma_1+3\gamma_2+p^2 \gamma_3 +5\gamma_4+p^2\gamma_5+G_{0*}+G_{1*})+(2m_\pi^2+q^2) \nonumber \\ && \times \Big( K_{0*}+2K_{1*} 
-(2m_\pi^2+q^2+4p^2){K_0+2K_1\over 4} \Big)\bigg]i(\vec\sigma_1\!+\!\vec\sigma_2)\!\cdot\!
(\vec q\!\times\!\vec p\,) \nonumber \\ && +3\Big[(2m_\pi^2+q^2+4p^2){K_2\over 2}-2K_{2*}-\gamma_2-G_2\Big] (q^2\vec\sigma_1\!\cdot\!\vec\sigma_2-\vec\sigma_1\!\cdot\! 
\vec q\, \vec\sigma_2\!\cdot\!  \vec q\,)  \nonumber \\ && +3 \Big[(2m_\pi^2+q^2+4p^2)\Big(
{K_0\over 2}+2K_1+2K_3\Big)-\gamma_0-2\gamma_1-\gamma_3 \nonumber \\ && -G_0-4G_1-4G_3-2(K_{0*}+4K_{1*}+4K_{3*})\Big] \vec\sigma_1\!\cdot\!(\vec q\!\times\!\vec p\,)\vec\sigma_2\!\cdot\!
(\vec q\!\times\!\vec p\,) \bigg\} \,.\end{eqnarray}
The new loop functions $K_\nu(p,q,k_f)$ are defined and given in explicit form in the appendix.
\subsection{$1/M$-correction to medium insertion}
One also needs to consider relativistic corrections to the in-medium N-propagator (or particle-hole propagator). Taking into account the (off-shell) kinetic energy $\vec l^{\,2}/2M$, the medium-insertion reads:
\begin{equation} -2\pi \delta\bigg(l_0-{\vec l^{\,2} \over 2M}\bigg) \theta(k_f-|\vec l\,|) 
= 2\pi \bigg[ -\delta(l_0)+{\vec l^{\,2} \over 2M}\delta'(l_0)+\dots \bigg] \theta(k_f-
|\vec l\,|) \,, 
\end{equation} 
where we have expanded in $1/M$. The $l_0$-integral is evaluated by the formula  $\int\!dl_0\, \delta'(l_0){\cal I}(l_0)= - {\cal I}'(0)$. When neglecting consistently the external kinetic energies $p_0=p_0' = p^2/2M$, the pion propagators as well as the isoscalar $c_{2,3}$-vertices 
\cite{3Nlong} give rise to an integrand ${\cal I}(l_0)$ that depends quadratically on $l_0$, thus  $ {\cal I}'(0)=0$. A linear dependence on $l_0$ is provided at leading order by the isovector Weinberg-Tomozawa $\pi\pi$NN-vertex. When combined with two ordinary $\pi$N-couplings, the pertinent diagram resembling a $2\pi$-exchange 3N-interaction generates (after closing one nucleon line to a loop) the following contributions to the in-medium NN-potential $V_\text{med}$:
\begin{equation} V_\text{med}^{(1)} = -{g_A^2 \over 32 \pi^2 M f_\pi^4}\,{ \vec\tau_1\!\cdot\! 
\vec\tau_2\over m_\pi^2+q^2}\big[ 3\Gamma_2+ 5\Gamma_4 +p^2(\Gamma_3+\Gamma_5)\big]
\vec\sigma_1\!\cdot\! \vec q\, \vec\sigma_2\!\cdot\! \vec q \,,  \end{equation} 
\begin{equation} V_\text{med}^{(3)} = {g_A^2 \vec\tau_1\!\cdot\! \vec\tau_2 \over 32 \pi^2 
M f_\pi^4}\Big\{6\Gamma_2+2p^2\Gamma_3-(2m_\pi^2+q^2)G_{0*} +(G_{0*}+2G_{1*})\,i (\vec\sigma_1
\!+\!\vec\sigma_2)\!\cdot\!( \vec q\!\times\!\vec p\,)\Big\} \,. \end{equation} 
Note that these contributions $V_\text{med}^{(1,3)}$ differ from those in eqs.(32,33), which had their origin in the $1/M$-correction to the Weinberg-Tomozawa vertex. 
\section{Subleading three-nucleon contact potential}
The subleading three-nucleon contact potential (appearing at N$^4$LO in the chiral counting) has been analyzed in ref.\cite{3Ncontact}. Taking into account Poincare symmetry and Fierz constraints the initial list of 116 operators could be reduced to 10 independent ones. The resulting 3N-contact potential $V_\text{3N}$, written in eq.(15) of ref.\cite{3Ncontact}, depends quadratically on the nucleon momenta and it involves ten parameters, called $E_1,\dots, E_{10}$.
Working out for each term $\sim E_i$ the closing of one nucleon line to a loop, one obtains from the subleading 3N-contact potential the following in-medium NN-interaction:
\begin{eqnarray} && {V_\text {med}\over \rho}=
E_1\Big( {6\over 5}k_f^2+2p^2-3q^2\Big)+E_2\Big[ (\vec \tau_1\!\cdot \!\vec
\tau_2+3)\Big({3\over 5}k_f^2+p^2\Big) -\vec \tau_1\!\cdot \!\vec \tau_2\, q^2 \Big]
\nonumber \\ && + E_3\Big[ (\vec \sigma_1\!\cdot \!\vec\sigma_2+3)\Big({3\over 5}k_f^2
 +p^2\Big) -\vec \sigma_1\!\cdot \!\vec \sigma_2\, q^2 \Big]+E_4\Big[(\vec \tau_1\!
\cdot \!\vec \tau_2\, \vec\sigma_1\!\cdot \!\vec\sigma_2+9)\Big({3\over 5}k_f^2+p^2\Big)
-\vec \tau_1\!\cdot \!\vec \tau_2\,\vec \sigma_1\!\cdot \!\vec \sigma_2\, q^2 \Big]
\nonumber \\ &&   +3 E_5\Big[ \vec \sigma_1\!\cdot\!\vec p\, \vec \sigma_2\!\cdot \!
\vec p+\vec \sigma_1\! \cdot \! \vec p\,' \vec \sigma_2\!\cdot \!\vec p\,'
+{2\over 5}k_f^2(\vec \sigma_1\!\cdot \!\vec \sigma_2-1) -{2\over 3} p^2+q^2-
\vec \sigma_1\!\cdot \!\vec q\, \vec \sigma_2\!\cdot \!\vec q\,\Big]\nonumber
\\ &&   +3 E_6\Big\{ (\vec \tau_1\!\cdot \!\vec \tau_2+3)\Big[{1\over 2}(\vec \sigma_1
\!\cdot\!\vec p\, \vec \sigma_2\!\cdot \!\vec p+\vec \sigma_1\! \cdot \! \vec p\,'  \vec
\sigma_2\!\cdot \!\vec p\,') -{p^2 \over 3}+{k_f^2\over 5}(\vec \sigma_1\!\cdot \!\vec \sigma_2
-1)\Big] +\vec \tau_1\!\cdot \!\vec \tau_2\Big( {q^2 \over 3}-\vec \sigma_1\!\cdot \!\vec q
\, \vec \sigma_2\!\cdot \!\vec q\Big)\Big\}\nonumber\\ &&  
+ (9E_8-7E_7) {i \over 2} (\vec \sigma_1\!+\!\vec \sigma_2)\!\cdot\!(\vec q\!\times\! \vec p\,) 
+E_9 \Big[  \vec \sigma_1\!\cdot \!\vec q\, \vec \sigma_2\!\cdot \!\vec q+{q^2\over 2}
-p^2-{3\over 5} k_f^2 -{i \over 2}(\vec \sigma_1\!+\!\vec \sigma_2)\!\cdot\!(\vec q\!\times\! \vec p\,)\Big]\nonumber\\ &&  + E_{10} \Big[ \vec \tau_1\!\cdot \!\vec \tau_2\, \vec \sigma_1\!
\cdot \!\vec q\, \vec \sigma_2\!\cdot \!\vec q+{3\over 2}q^2-3p^2-{9\over 5} k_f^2-{3i \over 2}
(\vec \sigma_1\!+\!\vec \sigma_2)\!\cdot\!(\vec q\!\times\! \vec p\,) \Big] \,,\end{eqnarray}
which depends linearly on the density $\rho= 2k_f^3/3\pi^2$ and quadratically on the momenta 
$\vec p, \vec q, k_f$. The above expression for $V_{\rm med}$ contributes only to  s- and p-wave matrix elements and to $^3\!S_1$-$^3\!D_1$ mixing.

\section*{Appendix: Loop-functions}
In this appendix we specify all the loop functions which have been used in section 5 to express 
the contributions to the in-medium NN-interaction $V_\text{med}$. 

The functions 
$\Gamma_\nu(p,k_f)$ with $\nu=0,\dots, 5$ arise from Fermi sphere integrals $(2\pi)^{-1}
\!\int\!d^3l\, \theta(k_f-|\vec l\,|)$ over a pion propagator $[m_\pi^2+(\vec l +\vec p\,)^2]^{-1}$, supplemented by tensorial factors $1\,(\nu=0), l_i\,(\nu=1), l_il_j\,(\nu=2,3)$ or $l_il_jl_k\,(\nu=4,5)$. Their respective analytical expressions read: 
\begin{equation}
\Gamma_0=k_f-m_\pi\Big[\arctan{k_f+p \over m_\pi}+\arctan{k_f-p\over m_\pi}\,\Big]+{m_\pi^2+ 
k_f^2-p^2\over 4p}\ln{m_\pi^2+(k_f+p)^2 \over m_\pi^2+(k_f-p)^2}\,, 
\end{equation}

\begin{equation}
\Gamma_1={k_f\over 4p^2}(m_\pi^2+k_f^2+p^2)-\Gamma_0-{1\over 16p^3}\big[m_\pi^2+(k_f+p)^2 
\big] \big[m_\pi^2+(k_f-p)^2]\ln{m_\pi^2+(k_f+p)^2 \over m_\pi^2+(k_f-p)^2}\,, 
\end{equation}

\begin{equation}
\Gamma_2={k_f^3\over 9}+{1\over 6}(k_f^2-m_\pi^2-p^2)\Gamma_0+{1\over 6}(m_\pi^2+k_f^2-p^2)\Gamma_{1}\,,
\end{equation}
\begin{equation}
\Gamma_3={k_f^3 \over 3p^2}-{m_\pi^2+k_f^2+p^2 \over 2p^2}\Gamma_0-{m_\pi^2+k_f^2+3p^2
\over 2p^2}\Gamma_1\,, 
\end{equation}

\begin{eqnarray}
\Gamma_4&=&{m_\pi^2\over 3}\Gamma_0  + {k_f \over 64}\bigg[{5p^2 \over 3}-3m_\pi^2-{31k_f^2 
\over 9}+{1 \over 3p^2}(3k_f^4-14k_f^2 m_\pi^2-17m_\pi^4)-{(k_f^2+m_\pi^2)^3 \over p^4}\bigg] \nonumber \\ && +{1\over 768p^5}\big[m_\pi^2+(k_f+p)^2 \big] \big[m_\pi^2
+(k_f-p)^2]\Big[3(k_f^2+m_\pi^2)^2+2p^2(k_f^2+7m_\pi^2)-5p^4\Big] \nonumber \\ && \times 
\ln{m_\pi^2+(k_f+p)^2 \over m_\pi^2+(k_f-p)^2}\,,
\end{eqnarray}

\begin{eqnarray}
\Gamma_5&=&-\Gamma_0  + {k_f \over 64}\bigg[29+{5 \over p^6}(k_f^2+m_\pi^2)^3+{25k_f^2+141 
m_\pi^2 \over 3p^2}+{1 \over 3p^4}(17k_f^4+86k_f^2 m_\pi^2+69m_\pi^4)\bigg] \nonumber \\ && -{1\over 256p^7}\big[m_\pi^2+(k_f+p)^2 \big] \big[m_\pi^2+(k_f-p)^2]\Big[5(k_f^2+m_\pi^2)^2 +2p^2(7k_f^2+9m_\pi^2)+29p^4\Big] \nonumber \\ && \times 
\ln{m_\pi^2+(k_f+p)^2 \over m_\pi^2+(k_f-p)^2}\,. \end{eqnarray}

The functions $\gamma_\nu(p,k_f)$ with $\nu=0,\dots, 5$ arise from Fermi sphere
integrals over a squared pion propagator $[m_\pi^2+(\vec l +\vec p\,)^2]^{-2}$, supplemented 
by tensorial factors $1\,(\nu=0), l_i\,(\nu=1), l_il_j\,(\nu=2,3)$ or 
$l_il_jl_k\,(\nu=4,5)$. They satisfy the relation $\gamma_\nu= - \partial \Gamma_\nu /\partial 
m_\pi^2$ and their analytical expressions read:
\begin{equation}
\gamma_0={1\over 2m_\pi}\Big[\arctan{k_f+p \over m_\pi}+\arctan{k_f-p\over m_\pi}\,\Big]-
{1\over 4p}\ln{m_\pi^2+(k_f+p)^2 \over m_\pi^2+(k_f-p)^2}\,, 
\end{equation}
\begin{equation}
\gamma_1=-\gamma_0-{k_f \over 2p^2}+{p^2+k_f^2+m_\pi^2\over 8p^3} \ln{m_\pi^2+(k_f+p)^2 
\over m_\pi^2+(k_f-p)^2}\,, 
\end{equation}
\begin{equation}
\gamma_2={k_f \over 8p^2}(3p^2-k_f^2-m_\pi^2)-m_\pi^2\gamma_0+{1\over 32p^3}\big[
(p^2+k_f^2+m_\pi^2)^2-4p^2(p^2+m_\pi^2)\big] \ln{m_\pi^2+(k_f+p)^2 \over m_\pi^2+(k_f-p)^2}\,, 
\end{equation}

\begin{eqnarray}
\gamma_3 &=&\gamma_0+{k_f \over 8p^4}(7p^2+3k_f^2+3m_\pi^2)-{1\over 32p^5}\big[ 3(k_f^2+
m_\pi^2)^2 \nonumber \\ &&+  2p^2(3k_f^2+5m_\pi^2)^2+7p^4\big] \ln{m_\pi^2+
(k_f+p)^2  \over m_\pi^2 +(k_f-p)^2}\,, \end{eqnarray}

\begin{eqnarray}
\gamma_4 &=&m_\pi^2\gamma_0+{k_f \over 16p^4}\Big[(k_f^2+m_\pi^2)^2+{4p^2 \over 3} (k_f^2+3
m_\pi^2)-5p^4\Big] \nonumber \\ &&+  {p^2-k_f^2-m_\pi^2 \over 64p^5} \big[5p^4+2p^2(k_f^2+3
m_\pi^2)+(k_f^2+m_\pi^2)^2\big] \ln{m_\pi^2+(k_f+p)^2  \over m_\pi^2 +(k_f-p)^2}\,, \end{eqnarray}

\begin{eqnarray}
\gamma_5 &=&-\gamma_0-{k_f \over p^2}\bigg[{5 \over 16p^4}(k_f^2+m_\pi^2)^2+{19 \over 16} 
+{2k_f^2+3 m_\pi^2 \over 3p^2}\bigg] +  {1 \over 64p^7} \big[5p^4 (3k_f^2+ 7m_\pi^2)
\nonumber \\ &&+19p^6 +3p^2(k_f^2+m_\pi^2)(3k_f^2+7m_\pi^2)+5(k_f^2+m_\pi^2)^3\big] \ln{m_\pi^2
+(k_f+p)^2  \over m_\pi^2 +(k_f-p)^2}\,. \end{eqnarray}

The functions $G_\nu(p,q,k_f)$ with $\nu=0,\dots, 3$ arise from Fermi sphere
integrals over two different pion propagators $[m_\pi^2+(\vec l +\vec p\,)^2]^{-1}[m_\pi^2+(\vec l +\vec p\,')^2]^{-1}$, multiplied by tensorial factors $1\,(\nu=0), l_i\,(\nu=1)$ or $l_il_j\,(\nu=2,3)$. In order to construct this set of functions one starts with the one-parameter (radial) integrals:
\begin{equation}
G_{0,0*,**}={2\over q}\int_0^{k_f}\!\!\!dl {\{l,l^3,l^5\}\over \sqrt{B+q^2 l^2}}\ln{q\,l+\sqrt{B+q^2l^2}\over \sqrt{B}}\,, \end{equation}
with the abbreviation $B=[m_\pi^2+(l+p)^2][m_\pi^2+(l-p)^2]$ and solves systems of linear equations:
\begin{eqnarray}
&& G_1={1\over 4p^2-q^2}\big[\Gamma_0-(m_\pi^2+p^2)G_0-G_{0*}\big]\,, \\
&& G_{1*}={1 \over 4p^2-q^2}\big[3\Gamma_2+p^2\Gamma_3-(m_\pi^2+p^2)G_{0*}-G_{**}\big]\,,\\
&& G_2=(m_\pi^2+p^2)G_1+G_{0*}+G_{1*}\,, \\
&& G_3={1\over 4p^2-q^2}\Big[{\Gamma_1\over 2}-2(m_\pi^2+p^2)G_1-G_{0*}-2G_{1*}\Big]\,. \end{eqnarray}

The functions $K_{\nu, \nu*}(p,q,k_f)$ with $\nu=0,\dots, 3$ arise from Fermi sphere
integrals over the symmetrized combination $[m_\pi^2+(\vec l +\vec p\,)^2]^{-2}[m_\pi^2+(\vec l 
+\vec p\,')^2]^{-1}+[m_\pi^2+(\vec l +\vec p\,)^2]^{-1}[m_\pi^2+(\vec l +\vec p\,')^2]^{-2}$ 
of pion propagators, supplemented by tensorial factors $1\,(\nu=0), l_i\,(\nu=1)$ or $l_il_j\,(\nu=2,3)$. By definition the relation $K_\nu= - \partial G_\nu /\partial m_\pi^2$ holds.
Again, one starts with four functions represented by one-parameter integrals:
\begin{equation}
K_{0,0*,**,***}=2\int_0^{k_f}\!\!\!dl\{l,l^3,l^5,l^7\}{m_\pi^2+l^2+p^2 \over B+q^2l^2} \bigg[{l \over B}+{1 \over q\sqrt{B+q^2l^2}}\ln{q\,l+\sqrt{B+q^2l^2} \over \sqrt{B}}\bigg]\,,
\end{equation}
and solves linear equations for the others:
\begin{eqnarray}
&& K_1={1\over 4p^2-q^2}\big[\gamma_0+G_0-(m_\pi^2+p^2)K_0-K_{0*}\big]\,, \\
&& K_{1*}={1 \over 4p^2-q^2}\big[3\gamma_2+p^2\gamma_3+G_{0*}-(m_\pi^2+p^2)K_{0*}-K_{**}\big]
\,,\\
&& K_2=(m_\pi^2+p^2)K_1-G_1+K_{0*}+K_{1*}\,, \\
&& K_3={1\over 4p^2-q^2}\Big[{\gamma_1\over 2}-2(m_\pi^2+p^2)K_1+2G_1-K_{0*}-2K_{1*}\Big]\,,\\
&& K_{1**}={1 \over 4p^2-q^2}\big[G_{**}-(m_\pi^2+p^2)K_{**}-K_{***}+\gamma_{**}\big] \,,\\
&& K_{2*}=(m_\pi^2+p^2)K_{1*}-G_{1*}+K_{**}+K_{1**}\,, \\
&& K_{3*}={1\over 4p^2-q^2}\Big[{1\over 2}(5\gamma_4+p^2\gamma_5)-2(m_\pi^2+p^2)K_{1*}+2G_{1*}
-K_{**}-2K_{1**}\Big]\,. \end{eqnarray}
We remind that a $*$ indicates an additional power of $\vec l\,^2$ in the integrand.
The new auxiliary function $\gamma_{**}$ in eq.(72) is given by:
\begin{equation}\gamma_{**}= (p^4+5m_\pi^4-10 p^2 m_\pi^2)\gamma_0 +4k_f\Big( p^2+{k_f^2 \over 6}
-m_\pi^2\Big) +{m_\pi^4 - p^4 \over p} \ln{m_\pi^2 +(k_f+p)^2  \over m_\pi^2 +(k_f-p)^2}\,. \end{equation}

\end{document}